# Electronic Raman scattering in Magnetite,

# Spin vs. Charge gap.

L.V. Gasparov, A. Rush.

*Department of Physics, University of North Florida,*

*1 UNF drive, Jacksonville, Fl 32224, USA*

G. Güntherodt

*2 Physikalisches Institut, RWTH-Aachen, 52056 Aachen, Germany*

H. Berger

*EPFL, CH-1015 Lausanne, Switzerland*

abstract

We report Raman scattering data of single crystals of magnetite ($Fe_3O_4$) with Verwey transition temperatures ($T_v$) of 123 and 117K, respectively. Both single crystals reveal broad electronic background extending up to 900 wavenumbers (~110 meV). Redistribution of this background is observed when samples are cooled below $T_v$. In particular, spectra of the low temperature phase show diminished background below 300 wavenumbers followed by an enhancement of the electronic background between 300 and 400 wavenumbers. To enhance the effect of this background redistribution we divide the spectra just below the transition by the spectra just above the transition. A resultant broad peak-like feature is observed, centered at 370 ±40 wavenumbers (45 ±5 meV). The peak position of this feature does not scale with the transition temperature. We discuss two alternative assignments of this feature to a spin or charge gap in magnetite.

# 1. INTRODUCTION

Magnetite ($Fe_3O_4$) is a naturally occurring mineral which is of interest to remarkably different fields of science. It is the first magnetic material known to mankind and it is the earliest compound known[1] to manifest a charge-ordering transition, discovered by Verwey in 1939. Magnetite is also an integral part of many living organisms. For instance, magnetotactic bacteria[2] and pigeons[3] use it for navigation along the Earth's magnetic field. Furthermore, it was reported that magnetite occurs in human brains and may play a role in the pathogenesis of the neurodegenerative diseases such as Alzheimer's.[4]

In condensed matter physics magnetite has recently attracted substantial attention[5,6] because its charge carriers exhibit strong spin polarization at the Fermi level. This compound has the potential to become one of the leading materials for spintronics. This has initiated an interest in high-quality films of magnetite on a semiconductor substrate. Such films could form the core element of a ferromagnet-semiconductor device.[6] Magnetite has been extensively studied for more than sixty years, yet the physics of this compound is not completely understood. Competition between electronic, lattice and magnetic degrees of freedom presents a substantial challenge in describing physics of magnetite. Difficulty in successful modeling of this iron oxide creates a nagging reminder for the scientific community as it tries to tackle such many element compounds as high temperature superconductors and colossal magneto resistance compounds.

Verwey transition in magnetite still remains an unsolved puzzle. At ambient pressure the Verwey transition of pure or near-stoichiometric magnetite is first-order. This transition occurs at $T_v$~123 K, with changes in crystal structure, latent heat, and a two-order of magnitude decrease in dc-conductivity. Oxygen deficiency or doping may reduce the transition temperature, cause the transition to become higher order, or suppress it completely. There are several competing models of the transition including Verwey and Hayaaman's[7] original order-disorder transition theory, Anderson's[8] long range order (LRO)-short range order (SRO) model, Cullen and Callen's[9] theory based on pure electron correlations, and polaron based theory of the transition.[10-13] However, none of these theories successfully describe the whole body of experimental data. Recently x-ray, neutron and electron diffraction experiments have cast considerable doubt on both Verwey and Hayaaman's order-disorder model and Anderson's LRO-SRO model.[14, 15] Recent photoemission studies[16,17] of magnetite indicate that there is no closing of the electronic band gap at the Verwey transition but rather shrinking of it by about 50 meV. The latter led to the statement that the Verwey transition is not actually metal to insulator transition but rather a semiconductor (or bad metal) to semiconductor transition.[16] In this context, questions naturally arise about the nature and value of the gap in this compound. Photoemission, with all its advantages is a surface sensitive technique[16]. Depending on the incoming photon energy it probes somewhere between 15 and 45 Å. That corresponds to not more than a dozen magnetite unit cells. In contrast, optical spectroscopy such as Raman and infrared spectroscopy probes to the depth of about 1000 Å. Furthermore resolution of the optical methods is about one order of magnitude better than that of photoemission. Table 1 summarizes available data for the gap between the upper edge of single particle density-of-states and chemical potential. Note that this gap corresponds to the half of the "optical" gap obtained from the infrared data

The optical band gap detected in the infrared experiment[18,19] may have two origins. It could be the opening of a charge gap but it can also be due to the decrease of carrier mobility because of polaron "condensation" at the Verwey transition. Unfortunately, it is difficult to separate these two processes.

Raman spectroscopy is a powerful technique capable of addressing the issue of the gap in the magnetite. Such a gap should be manifested in the electronic background of magnetite's Raman spectrum as low intensity broad peaks. One could expect a diminished background at the frequencies smaller than the gap with subsequent increase of the scattering. Similar behavior has been observed in Raman studies of HTSC compounds.[20] Motivated by this fact we looked for potential signatures of the gap opening in the Raman spectra of $Fe_3O_4$.

In this paper we present new results on electronic Raman scattering obtained on two samples of magnetite grown by two different groups. We believe that we observe "gap" feature in our specta. We discuss the origins of this gap, particularly addressing two competing explanations. We conclude that this gap feature we observe has magnetic origin.

## 2. EXPERIMENTAL DETAILS

In this work we used two magnetite crystals from two different sources. The first crystal was grown by Dr. Berger in EPFL Lausanne. This was the same crystal used for our previous measurements in magnetite.[19,21] This crystal was grown by a chemical vapor transport technique using stoichiometric $Fe_3O_4$ microcrystalline powder obtained by reduction reaction of ferric oxide ($Fe_2O_3$). This procedure yielded near-stoichiometric single crystals with typical size of 4 x 4 x 1 mm. X-ray diffraction confirmed the spinel-type structure of the crystals. Transport measurements detect drop of conductivity at $T_v$ =123 K.

The second crystal with $T_V$=117K was provided by Dr. Pimenov from the University of Wurzburg. This crystal was originally grown by Dr. Brabers and crystals from the same batch were used for the Photoemission measurements of Ref. 16

Raman measurements were carried out on freshly cleaved surfaces of the as-grown single crystals. The 532 nm line of a CW Solid State laser as well as 514nm and 488 nm lines of an $Ar^+$ ion laser were used as excitation with no more than 10 mW incident power on the sample, in order to avoid oxidation.[22] Estimated overheating of the sample did not exceed 13K. Polarized Raman spectra were measured using a Dilor XY modular triple spectrometer equipped with a liquid nitrogen cooled CCD detector. The spectra were measured in nearly back scattering geometry. The sample temperature was maintained in He-bath cryostat over a range of 4.2 to 300K.

## 3. RESULTS AND DISCUSSION

The effects of the Verwey transition on the Raman spectra of magnetite have been addressed in a number of publications.[19,21-27] Below the Verwey transition magnetite has a much bigger unit cell[14] that leads to a dramatic increase in the number of phonon modes. In addition to the phonon modes, magnetite displays a rather broad electronic background extending up to 900 cm$^{-1}$. Zero momentum electronic and magnetic excitations in magnetite are likely source of this background.

This paper focuses on the evolution of this background above and below the transition. Fig. 1 displays a series of unpolarized spectra of magnetite near the Verwey transition. One can clearly see the enhancement of low frequency background in the high temperature phase. This effect is observed for both samples of magnetite in Figs. 1 (a) and 1 (b). To understand this enhancement we need to note that the Stokes Raman scattering cross section is directly proportional to the imaginary part of the Raman response function χ multiplied by the Bose-factor:

$$\frac{d^2\sigma}{d\Omega d\omega} \propto \left\{\frac{1}{1-e^{-\frac{\hbar\omega}{k_B T}}}\right\} \chi''(\omega,T) \qquad (1)$$

The imaginary part of the response function contains all the relevant information about quasi particle excitations in the sample under investigation. The observed background enhancement in the high temperature phase cannot be accounted by the Bose-factor contribution alone since equation (1) would be virtually unchanged over the small temperature shift at the Verwey transition.

To understand polarization dependence of this enhancement we measured polarized spectra of magnetite. Fig.2 and 3 display Raman data in the xx-polarization. In both figures we display Raman response function obtained by dividing corresponding spectra by the Bose factor. Overheating of 13K was taken into account. Spectrum below the Verwey transition (solid line) demonstrates redistribution of the electronic background as compared to the spectrum above the transition (dotted line). There is an enhancement of the background above 300 wavenumbers and clear depletion of the background below 400 wavenumbers.

To underline the redistribution of the background one can look at the frequency range between 400 and 500 wavenumbers. This spectral range has a single phonon mode at 470 cm$^{-1}$ that is absent above the Verwey transition. For this frequency range both Fig.2 and 3 indicate that the background below transition is substantially smaller than that above the transition. All this in spite of the fact that additional phonon mode appears in this spectral range below the transition temperature. So even though there is added spectral weight due to the phonon mode the background remains diminished compared to the above transition spectrum. We believe that such behavior is consistent with the opening of the gap below 375 wavenumbers.

In order to get a better sense of the background redistribution we divide the response function below the transition temperature by that above the transition. The net result is a peak-like feature with the maximum around 350 wavenumbers (Figs. 2 & 3, insets). We assign the peak position of that feature to the value of the gap. Clearly, strong phonon modes in the low temperature phase do not allow for precise determination of the gap value, however we believe that the manifestation of the gap opening is evident. Varying the excitation line frequency of our Ar$^+$ laser did not significantly reduce the phonon mode intensities. Our estimate of the uncertainty of this gap value is 40 wavenumbers (~5 meV). This feature is observed only in the XX-polarization. Corresponding XY spectra did not display any enhancement of the background.

One observes a number of new phonon modes in the low temperature phase. It is interesting to note that the modes below 300 cm$^{-1}$ have smaller line width compared to the modes above this frequency, Fig.4. This could be additional indication of strong interaction of the phonon modes with the opening of the gap

The value of the gap obtained from our experiment is at odds with that from the infrared data. One can make following suggestions to reconcile this differences. i) The infrared gap[18,19] value may be affected by the increase of carrier mobility associated with the polaron "condensation" at the Verwey transition; ii) The infrared gap is averaged over whole Brillouin zone (BZ) whereas Raman scattering displays a k-dependent charge gap. Photoemission data do indicate strong dependence of the photoemission gap on the crystallographic direction.[16]

Having made these statements, we recognize that differences between the values of the Raman and photoemission/ir gap are nearly a factor of two which may be too much to be accounted by the potential anisotropy of the charge gap. Therefore an alternative explanation of our data is in order.

Such explanation comes from the recent neutron scattering experiments of McQueeney et al.[28] In this work several acoustic and optical spin wave modes (magnons) were studied. Some of these magnetic excitations have been originally measured in the seminal neutron scattering experiment of Brockhouse,[29] however never detected by optical measurements. McQueeney's neutron data[28] indicates one particular acoustic magnon that is very sensitive to the Verwey transition. Above transition neutron scattering detects this magnon as a peak at (4,0, -1/2) reciprocal space extending from 30 to 50 meV (roughly 240-500 wavenumbers). The magnon peak itself demonstrates isotropic dispersion around the (004) Brillouin zone center.[28] Below the Verwey transition this peak splits into two peaks. The latter has been interpreted as an opening of a gap in the acoustic magnon branch. In the reciprocal space the gap is localized at ±0.1-0.2 reciprocal lattice units[28] along (0.0,1) direction. This gap is opening at 43 meV at the wave vector **q**= (0, 0, ½). This value of the "spin wave gap" is in excellent agreement with our data. Raman scattering probes zone center excitations in Solids. Acoustic magnon has zero energy at zero momentum and therefore it can not be detected through conventional Raman process. However two magnon scattering may alleviate this restriction. The mechanism of this two magnon scattering assumes interaction of two magnons with opposite q-vectors. Combined excitation will therefore have zero net momentum but non zero energy. Such excitation is averaged over Brillouin Zone and can be probed by appropriate Raman vertex. Based on this assumption we would expect two magnon Raman scattering to manifested itself as a broad peak.[30,31] Furthermore when gap develops in the magnon branch we would expect a drop in the intensity of these two magnon peak beyond the energy of the gap. XX Raman vertex probes over whole BZ. Furthermore the location of the gap in the magnon branch is consistent with this geometry. Since gap develops at substantially lower temperature we can also expect narrowing of the two-magnon peak.

In addition to magnetic excitations one can expect electronic excitations to play important role in the magnetite. In general one should see electronic excitations associated with carriers. The scattering rate[32,33] is related to the slope of the imaginary part of the Raman response function $\frac{\partial \chi''}{\partial \omega}$ in the limit of ω→0. The smaller the slope the shorter is the lifetime τ and the larger is the scattering rate Γ=1/τ. Resistivity is directly proportional to the scattering rate. When magnetite undergoes the Verwey transition its resistivity increases by about two orders of magnitude. The latter means that the carrier scattering rate should exhibit a jump at the transition. That means that the low temperature spectrum of magnetite should display low frequency scattering characterized by substantially flatter slope. That is indeed the case in our experiment.

This "spin gap" may be associated with the optical phonon mode crossing the acoustic magnon branch. McQueeney et al.[28] indicated possible flattening of the magnon branch at the opening of the gap. Such flattening may create a peak like singularity in Raman data, as we observe in our data.

Additional evidence for the spin gap nature of the observed Raman effects may come from the phonon line widths displayed in Fig.4. Below the transition in both samples the phonon modes become substantially broader above 370-380 cm$^{-1}$. This is likely an indication of strong coupling between magnetic and lattice excitations (hence XX polarization of the "spin gap" effects[34]), providing additional evidence of the spin gap. One should keep in mind that most of these modes are associated with the same Fe-O tetrahedron; therefore the increase may not be interpreted as merely different groups of ions that are somehow more affected by disorder. Increase of the line width is likely an indicator of additional decay channel for the phonons.

The XX geometry above the transition displays just three phonon modes, Fig. 1-2. These modes are centered at 310, 545 and 670 cm$^{-1}$ (Fig.1, 2). The widths of the modes at 150K are 60, 29, and 63 wavenumbers, respectively. Clearly these numbers do not follow the low temperature trend displayed in Fig.4. Only below the transition we observe a broadening of the modes centered above 370 wavenumbers. This clearly dovetails the idea of spin gap opening effect on the phonon modes width. It is therefore plausible to assign the 45 meV feature we observed in the Raman spectrum to the "spin wave gap".

This assignment, however reasonable, assumes that the charge gap is absent in Raman. This is probably due to small matrix elements for these particular excitations. We are not aware of any theoretical work addressing the issue of Raman tensors and the charge gap in magnetite. Our result points out the need for such calculations. Further experiments with Fe ions substituted by nonmagnetic ions will shed more light on to this issue.

## 4. CONCLUSION

In conclusion we present the Raman measurements of the electronic background in magnetite. Comparison of the spectra above and below the Verwey transition yields the gap value of the order of $45 \pm 5$ meV which is in excellent agreement with recent neutron scattering data on the spin wave gap at 43 meV.

## ACKNOWLEDGEMENTS

We would like to thank V.A.M. Brabers from the Eindhoven University of Technology and A. Pimenov from the University of Wurzburg for one of the single crystals of magnetite.

This work was supported by the NSF DMR-0805073 award, Alexander von Humboldt Foundation, Research Corporation Cottrell College Science award No. CC 6130, Petroleum Research Fund award No. 40926-GB10 and ONR N00014-06-1-0133 award.

- 13 -

**Tables**

Table 1. The value of the gap obtained from the photoemission and optical measurements of Magnetite.

| Measurement Method | Gap above the Verwey transition | Gap below the Verwey transition |
|---|---|---|
| Photoemission[16] | 40 meV | 90 meV |
| Photoemission[17] | 100 meV | 150 meV |
| Infrared[18] |  | 70 meV |
| Infrared[19] |  | 100 meV |

**Figure captions**

Fig.1 Unpolarized Raman spectra of magnetite in the XX geometry near Verwey transition in the (a) 123K- and (b)117K-samples. Spectra are taken with temperature increments of 4 K. Dashed line indicates the spectra above the Verwey transition and solid line indicates the spectra below the Verwey transition. Estimated overheating was around 13K. Note new phonon modes below the transition and strong redistribution of the electronic background as sample undergoes the transition.

Fig.2 Polarized (XX geometry) Raman spectra of magnetite with $T_v$ =123K. Shown are spectra above the transition at 125K (dashed line) and below the transition at 50K (solid line). Inset displays the ratio of the 50K-spectrum to that above the transition at 125K. We assign a broad peak-like feature around 350 wavenumbers to the opening of a gap. Horizontal dotted line at the spectra ration equal to one is a guide to the eye to appreciate the redistribution of the background.

Fig.3 Polarized (XX geometry) Raman spectra of magnetite with $T_v$ =117K. Shown are spectra above the transition at 125K (dashed line) and below the transition at 50K (solid line). Inset displays the ratio of the 50K-spectrum to that above the transition at 125K. We assign a broad peak-like feature around 350 wavenumbers to the opening of a gap. Horizontal dotted line at the spectra ration equal to one is a guide to the eye to appreciate the redistribution of the background.

Fig.4 The line width of the phonon modes in the 123K-(open diamonds) and 117K-(open circles) samples as a function of their frequency as measured at 50K (below the Verwey transition). The line width was obtained from the fit to the Lorenzian line shape. Dotted line is a guide to the eye. Strong increase of the line width for the phonon modes above 370 cm$^{-1}$ is evident. The increase is observed in both magnetite samples.

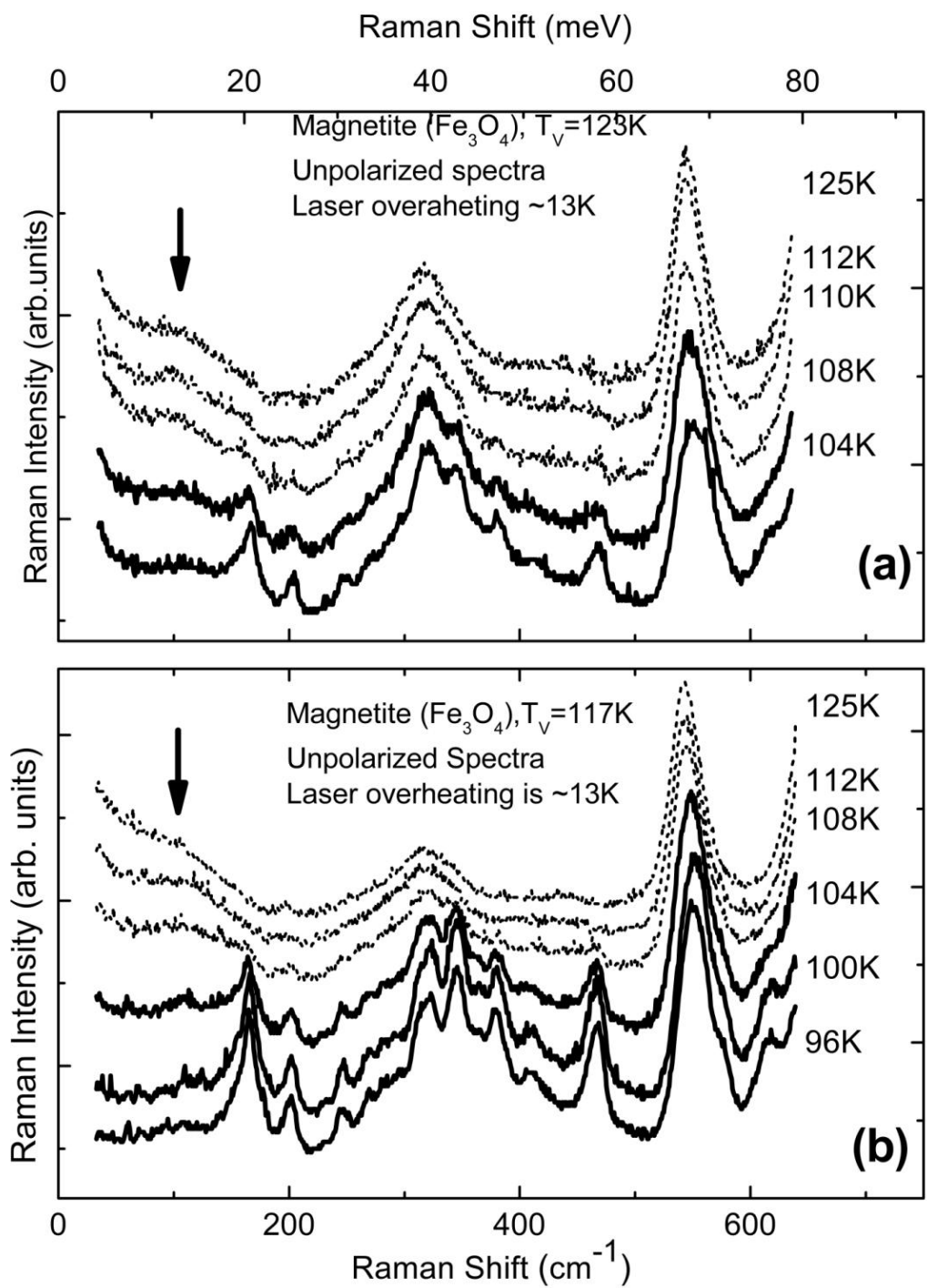

Fig.1 a,b

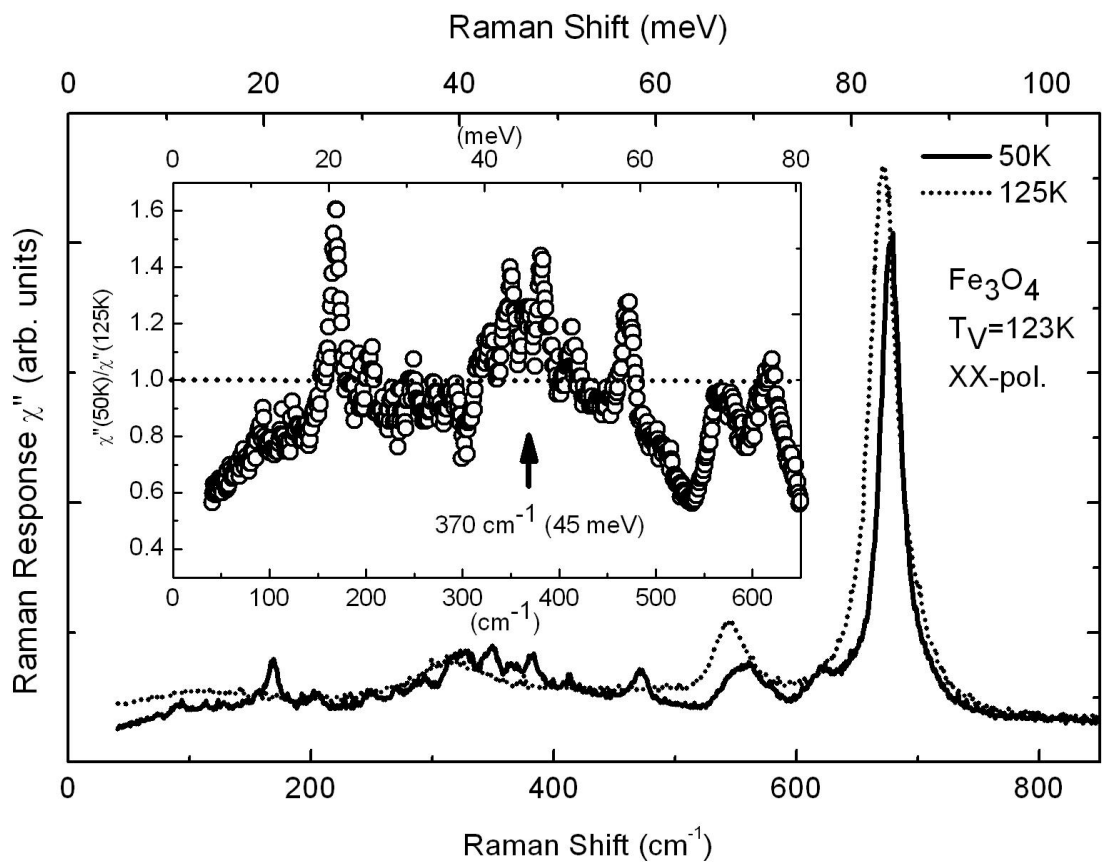

Fig.2

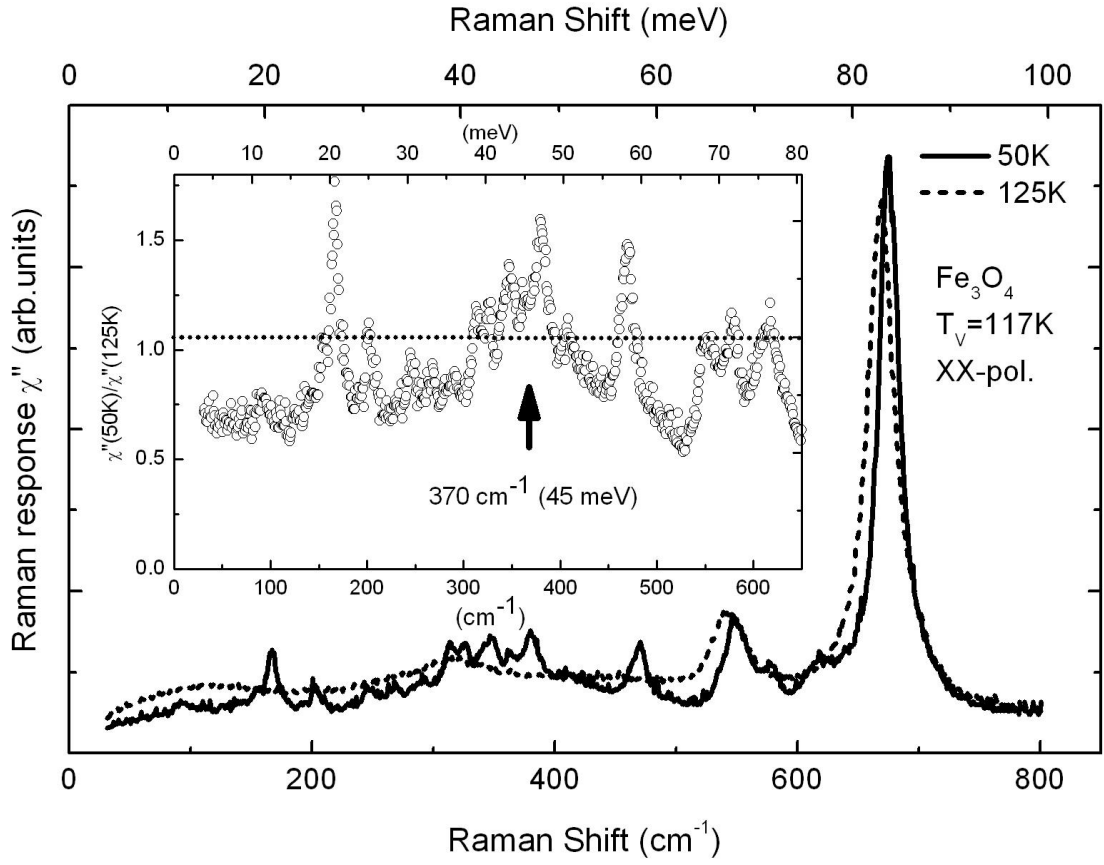

Fig.3

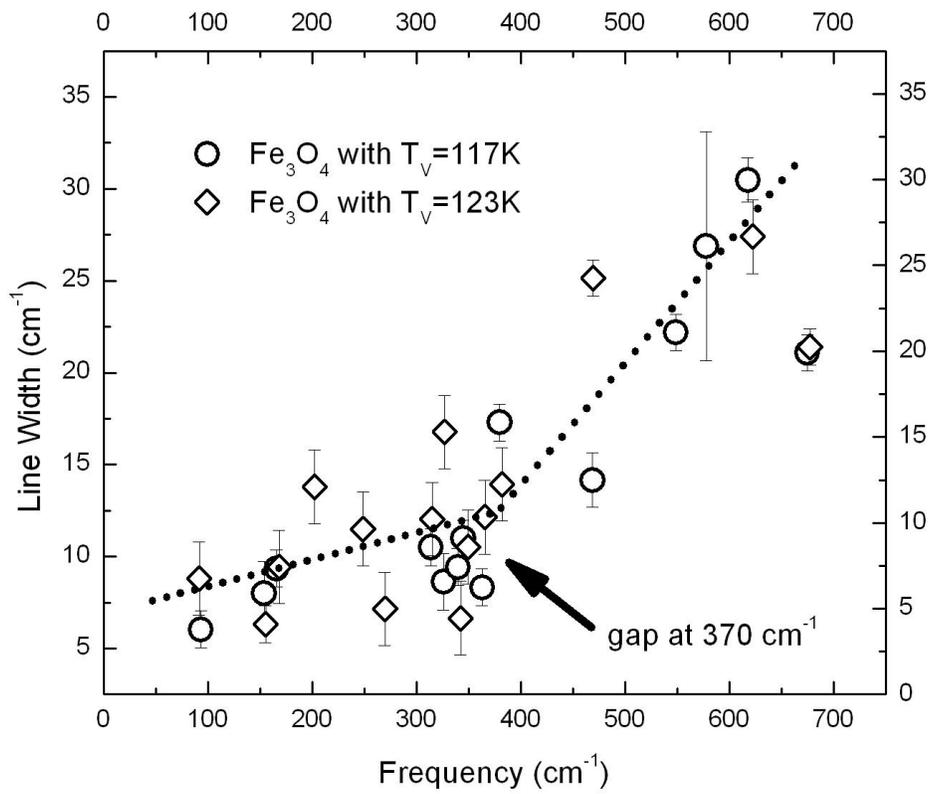

Fig.4